\documentclass[12pt,reqno]{amsart} % add titlepage param for separate title page
\usepackage[foot]{amsaddr}
\usepackage[utf8]{inputenc}
\usepackage[english]{babel}
\usepackage[shortlabels]{enumitem}

\usepackage{
  amsbsy,
  amscd,
  amsfonts,
  amsmath,
  amssymb,
  amsthm,
  bm, % boldmath
  booktabs,
  braket,
  cancel, % cancellation
  caption,
  cases,
  dsfont,
  epsfig,
  etoolbox,
  float,
  graphicx,
  hyperref,
  latexsym,
  lineno,
  lipsum,
  mathrsfs,
  mathtools,
  setspace,
  subfig,
  url,
  xargs,
  xcolor
}

\usepackage{array}   % for \newcolumntype macro
\newcolumntype{L}{>{$}l<{$}} % math-mode version of "l" column type

\makeatletter
\def\paragraph{\@startsection{paragraph}{4}%
  \z@\z@{-\fontdimen2\font}%
  {\normalfont\bfseries}}
\makeatother

\usepackage[ruled, lined, linesnumbered, commentsnumbered, longend]{algorithm2e}

\makeatletter
\def\@setthanks{\vspace{-\baselineskip}\def\thanks##1{\@par##1\@addpunct.}\thankses}
\makeatother

\usepackage{lscape}
\newcommand{\bland}{\begin{landscape}}
\newcommand{\eland}{\end{landscape}}

\newcommand{\bburl}[1]{\textcolor{blue}{\url{#1}}}

\definecolor{maroon}{rgb}{0.5, 0.0, 0.0}

\hypersetup{breaklinks=true,
            bookmarks=true,
            pdfauthor={Apoorva Lal},
             pdfkeywords = {},
            colorlinks=true,
            citecolor=maroon,
            urlcolor=blue,
            linkcolor=blue,
            pdfborder={0 0 0}}
\makeatletter
\def\maxwidth{\ifdim\Gin@nat@width>\linewidth\linewidth\else\Gin@nat@width\fi}
\def\maxheight{\ifdim\Gin@nat@height>\textheight\textheight\else\Gin@nat@height\fi}
\makeatother
\setkeys{Gin}{width=\maxwidth,height=\maxheight,keepaspectratio}

\newcommand{\burl}[1]{\textcolor{blue}{\url{#1}}}

\numberwithin{equation}{section}

\usepackage[colorinlistoftodos,prependcaption,]{todonotes}
\makeatletter
\providecommand\@dotsep{5}
\def\listtodoname{List of Todos}
\def\listoftodos{\@starttoc{tdo}\listtodoname}
\makeatother

\SetKwInput{kwParam}{Parameter}
\SetKwInput{kwInit}{Param}

\SetCommentSty{mycommfont}

\newcommand{\beq}{\begin{equation}}
\newcommand{\eeq}{\end{equation}}

\newcommand*\Bigpar[1]{\left( #1 \right )}

\newcommand*\SetB[1]{\left\{ #1 \right\}}
\newcommand*\Sett[1]{\mathcal{#1}}
\newcommand{\ba}{\begin{array}}
\newcommand{\ea}{\end{array}}
\newcommand{\be}{\begin{enumerate}}
\newcommand{\ee}{\end{enumerate}}
\newcommand{\bi}{\begin{itemize}}
\newcommand{\ei}{\end{itemize}}
\newcommand{\I}{\item}
\newcommand{\bs}{\begin{align}\begin{split}\nonumber}
\newcommand{\bsnumber}{\begin{align}\begin{split}}
\newcommand{\es}{\end{split}\end{align}}

\def\AND{\text{ and }}

\newcommandx{\deriv}[2][1=x,2=f]{\nabla \, #2 \Bigpar{ #1 } }
\newcommandx{\ortho}[1][1=L]{#1^{\bot}}

\newcommandx*\seqq[3][1=1,2=x, 3=n]{#2_{#1},\ldots,#2_{#3}}
\newcommandx*\coord[3][1=1,2=x, 3=n]{(#2_{#1},\ldots,#2_{#3})}

\newcommand{\sumin}{\ensuremath{\sum_{i=1}^n}}

\newcommand*\Indic[1]{\mathds{1}_{#1}}

\newcommand{\abs}[1]{\left\vert {#1} \right\vert}
\newcommand{\R}{\ensuremath{\mathbb{R}}}

\newcommand\frakfamily{\usefont{U}{yfrak}{m}{n}}
\DeclareTextFontCommand{\textfrak}{\frakfamily}

\newcommand{\ooN}{\frac{1}{n}}  %oneforth
\def\mbi#1{\boldsymbol{#1}} % Bold and italic (math bold italic)
\def\ve#1{\mbi{#1}} % Vector notation
\def\vee#1{\mathbf{#1}} % Vector notation
\newcommand{\wh}[1]{\widehat{#1}} % Wide hat
\newcommand{\E}{\mathbb{E}} % Expectation symbol
\providecommand{\abs}{\mathop{\rm abs}}

\definecolor{ao}{rgb}{0.0, 0.5, 0.0}
\definecolor{purp}{HTML}{5601A4}
\definecolor{navy}{HTML}{0D3D56}
\definecolor{ruby}{HTML}{9a2515}
\definecolor{corn}{HTML}{107895}
\definecolor{daisy}{HTML}{EBC944}
\definecolor{coral}{HTML}{F26D21}
\definecolor{kelly}{HTML}{829356}
\definecolor{cranb}{HTML}{E64173}
\definecolor{jet}{HTML}{131516}
\definecolor{ash}{HTML}{555F61}
\definecolor{slate}{HTML}{314F4F}

\newcommand\indep{\protect\mathpalette{\protect\independenT}{\perp}}
\def\independenT#1#2{\mathrel{\rlap{$#1#2$}\mkern5mu{#1#2}}}

\newcommand{\Exp}[1]{\mathbb{E}\left[#1\right]}
\newcommand{\Prob}[1]{\mathbf{Pr}\left(#1\right)}
\newcommand{\hyp}[2]{
\ensuremath{H_0:#1 \ifhmode\quad\text{versus}\quad\fi\text{ vs. } H_1:#2}}

\newcommandx{\uniff}[1][1={a,b}]{\textrm{Unif}\left({#1}\right)}
\newcommandx{\unifd}[1][1={a,\ldots,b}]{\textrm{Unif}\left\{{#1}\right\}}
\newcommandx{\dunif}[3][1=x,2=a,3=b]{\frac{I(#2<#1<#3)}{#3-#2}}
\newcommandx{\dunifd}[3][1=x,2=a,3=b]{\frac{I(#2\le#1\le#3)}{#3-#2+1}}
\newcommandx{\punif}[3][1=x,2=a,3=b]{
\begin{cases} 0 & #1 < #2 \\ \frac{#1-#2}{#3-#2} & #2 < #1 < #3 \\ 1 & #1 > #3\\\end{cases}}
\newcommandx{\punifd}[3][1=x,2=a,3=b]{
\begin{cases} 0 & #1 < #2\\ \frac{\lfloor#1\rfloor-#2+1}{#3-#2} & #2 \le #1 \le #3 \\ 1 & #1 > #3\\ \end{cases}}

\newcommandx\bern[1][1=p]{\textrm{Bern}\left({#1}\right)}
\newcommandx\dbern[2][1=x,2=p]{#2^{#1} \left(1-#2\right)^{1-#1}}
\newcommandx\pbern[2][1=x,2=p]{\left(1-#2\right)^{1-#1}}

\newcommandx\bin[1][1={n,p}]{\textrm{Bin}\left(#1\right)}
\newcommandx\dbin[3][1=x,2=n,3=p]{\binom{#2}{#1}#3^#1\left(1-#3\right)^{#2-#1}}

\newcommandx\mult[1][1={n,p}]{\textrm{Mult}\left(#1\right)}
\newcommandx\dmult[3][1=x,2=n,3=p]{\frac{#2!}{#1_1!\ldots#1_k!}#3_1^{#1_1}\cdots#3_k^{#1_k}}

\newcommandx\hyper[1][1={N,m,n}]{\textrm{Hyp}\left({#1}\right)}
\newcommandx\dhyper[4][1=x,2=N,3=m,4=n]{\frac{\binom{#3}{#1}\binom{#2-#3}{#4-#1}}{\binom{#2}{#4}}}

\newcommandx\nbin[1][1={r,p}]{\textrm{NBin}\left({#1}\right)}
\newcommandx\dnbin[3][1=x,2=r,3=p]{\binom{#1+#2-1}{#2-1}#3^#2(1-#3)^#1}
\newcommandx\pnbin[3][1=x,2=r,3=p]{I_#3(#2,#1+1)}

\newcommandx\geo[1][1=p]{\textrm{Geo}\left(#1\right)}
\newcommandx\dgeo[2][1=x,2=p]{#2(1-#2)^{#1-1}}
\newcommandx\pgeo[2][1=x,2=p]{1-(1-#2)^#1}

\newcommandx\pois[1][1=\lambda]{\textrm{Po}\left({#1}\right)}
\newcommandx\dpois[2][1=x,2=\lambda]{\frac{#2^#1 e^{-#2}}{#1!}}
\newcommandx\ppois[2][1=x,2=\lambda]{e^{-#2}\sum_{i=0}^#1\frac{#2^i}{i!}}

\newcommandx\normall[1][1={\mu,\sigma^2}]{\mathcal{N}\left({#1}\right)}
\newcommandx\dnormall[3][1=x,2=\mu,3=\sigma]%
  {\frac{1}{#3\sqrt{2\pi}}\exp \Bigpar{-\frac{\left(#1-#2\right)^2}{2 #3^2}}}
\newcommandx\pnormall[1][1=x]{\Phi\left({#1}\right)}
\newcommandx\qnormall[1]{\Phi^{-1}\left({#1}\right)}

\newcommandx\mvn[1][1={\mu,\Sigma}]{\mathrm{MVN}\left({#1}\right)}

\newcommandx\ex[1][1=\lambda]{\textrm{Exp}\left(#1\right)}
\newcommandx\dex[2][1=x,2=\lambda]{#2e^{-#1 #2}}
\newcommandx\pex[2][1=x,2=\lambda]{1-e^{-#1 #2}}

\newcommandx\gam[1][1={\alpha,\lambda}]{\textrm{Gamma}\left({#1}\right)}
\newcommandx\dgamma[3][1=x,2=\alpha,3=\lambda]%
  {\frac{#3^{#2}}{\Gamma\left( #2 \right)} #1^{#2-1}e^{-#3#1}}

\newcommandx\invgamma[1][1={\alpha,\beta}]{\textrm{InvGamma}\left({#1}\right)}
\newcommandx\dinvgamma[3][1=x,2=\alpha,3=\beta]%
{\frac{#3^{#2}}{\Gamma\left(#2\right)}#1^{-#2-1}e^{-#3/#1}}
\newcommandx\pinvgamma[3][1=x,2=\alpha,3=\beta]%
{\frac{\Gamma\left(#2,\frac{#3}{#1}\right)}{\Gamma\left(#2\right)}}

\newcommandx\bet[1][1={\alpha,\beta}]{\textrm{Beta}\left(#1\right)}
\newcommandx\dbeta[3][1=x,2=\alpha,3=\beta]
{\frac{\Gamma\left(#2+#3\right)}{\Gamma\left(#2\right)\Gamma\left(#3\right)}#1^{#2-1}\left(1-#1\right)^{#3-1}}

\newcommandx\dir[1][1={\alpha}]{\textrm{Dir}\left(#1\right)}
\newcommandx\ddir[3][1=x,2=\alpha]{\frac{\Gamma\left(\sum_{i=1}^k #2_i\right)}{\prod_{i=1}^k\Gamma\left(#2_i\right)}\prod_{i=1}^k #1_i^{#2_i-1}}

\newcommandx\weibull[1][1={\alpha}]{\textrm{Dir}\left(#1\right)}
\newcommandx\dweibull[3][1=x,2=\lambda,3=k]{\frac{#3}{#2}
\left(\frac{#1}{#2}\right)^{#3-1} e^{-(#1/#2)^k}}

\newcommandx\chisq[1][1=k]{\chi_{#1}^2}

\newcommandx\zet[1][1=s]{\textrm{Zeta}\left(#1\right)}
\newcommandx\dzeta[2][1=x,2=s]{\frac{#1^{-#2}}{\zeta\left(#2\right)}}

\makeatletter%
  \@ifclassloaded{beamer}%
  {}% skip all below
  { % else
    \newtheoremstyle{mystyle}%                % Name
      {}%                                     % Space above
      {}%                                     % Space below
      {}%                                     % Body font
      {}%                                     % Indent amount
      {\sffamily \bfseries }%                            % Theorem head font
      {.}%                                    % Punctuation after theorem head
      {\newline }%                                    % Space after theorem head, ' ', or \newline
      {\thmname{#1}\thmnumber{ #2}\thmnote{ (#3)}}%                                     % Theorem head spec (can be left empty, meaning `normal')
    \theoremstyle{mystyle}

    \newenvironment{proof-sketch}{\noindent{\bf Sketch of Proof}
      \hspace*{1em}}{\qed\bigskip\\}
    \newenvironment{proof-idea}{\noindent{\bf Proof Idea}
      \hspace*{1em}}{\qed\bigskip\\}
    \newenvironment{proof-of-lemma}[1][{}]{\noindent{\bf Proof of Lemma {#1}}
      \hspace*{1em}}{\qed\bigskip\\}
    \newenvironment{proof-of-proposition}[1][{}]{\noindent{\bf
        Proof of Proposition {#1}}
      \hspace*{1em}}{\qed\bigskip\\}
    \newenvironment{proof-of-theorem}[1][{}]{\noindent{\bf Proof of Theorem {#1}}
      \hspace*{1em}}{\qed\bigskip\\}
    \newenvironment{inner-proof}{\noindent{\bf Proof}\hspace{1em}}{
      $\bigtriangledown$\medskip\\}
    \newenvironment{proof-attempt}{\noindent{\bf Proof Attempt}
      \hspace*{1em}}{\qed\bigskip\\}

  }%
\makeatother%

\usepackage[margin=1in]{geometry}
\usepackage{makecell}

\usepackage[width=.9\textwidth]{caption}

\usepackage{enumitem}
\setlist[enumerate]{noitemsep, topsep=0pt}

\usepackage[
    backend=biber,
    ]{biblatex}
\renewbibmacro{in:}{}
\begin{document}

\title[Framework for Causal Bridging]{a framework for generalization and transportation of causal estimates under covariate shift}

\author{Apoorva Lal} 
\email{{\href{apoorval@stanford.edu}{apoorval@stanford.edu}}}
% \address{Stanford, Netflix DSE}

\author{Wenjing Zheng} 
\email{{\href{mailto:wzheng@netflix.com}{wzheng@netflix.com}}}
% \address{Netflix DSE}

\author{Simon Ejdemyr}
\email{{\href{mailto:sejdemyr@netflix.com}{sejdemyr@netflix.com}}}
% \address{Netflix DSE}

\keywords{experimentation, generalization, transportation, bridging}
\date{\today}

\maketitle

% \section*{Introduction} \label{intro}
%%%%%%%%%%%%%%%%%%%%%%%%%%%%%%%%%%%%%%%%%%%%%%%%%%%%%%%%%%%%

Randomized experiments are an excellent tool for estimating
\emph{internally valid} causal effects with the sample at hand, but
their \emph{external} validity is frequently questioned. While
classical results on the estimation of Population Average Treatment
Effects (PATE) implicitly assume random selection into experiments,
this is typically far from true in many medical, social-scientific,
and industry experiments. When the experimental sample is different
from the target sample along observable or unobservable dimensions
(termed \emph{covariate shift} in the causal learning literature),
experimental estimates may be of limited use for policy decisions. We
cast this as a sample selection problem and propose methods to
re-weight the doubly-robust scores from experimental subjects to
estimate treatment effects in the overall sample (=:
\emph{generalization}) or in an alternate target sample (=:
\emph{transportation}). We implement these estimators in the
open-source package
\href{https://github.com/Netflix-Skunkworks/causalTransportR}{\texttt{causalTransportR}}\footnote{Available
at
\href{https://github.com/Netflix-Skunkworks/causalTransportR}{https://github.com/Netflix-Skunkworks/causalTransportR}}
and illustrate its performance in a simulation study and discuss
diagnostics to evaluate its performance.

\vspace{-6mm}
\section*{Methods}
\vspace{-6mm}

We observe $n$ iid copies of $(\vee{X}_i, S_i, S_i A_i, S_i Y_i)_{i =
1}^n$, where covariates $\vee{X}_i \in \R^p$, treatment $A_i \in
\Sett{A} := \SetB{0, \dots, K}$, outcome $Y_i \in \R$, and selection
indicator $S_i \in \SetB{0, 1}$ is a function of pre-treatment
variables and is not affected by treatment. In other words, we observe
$(\vee{X}_i, A_i, Y_i)_{i = 1}^{N_1}$ for observations with $S_i = 1$
(henceforth the {\it study} sample $\Sett{S}_1$), and only
$(\vee{X}_i)_{i = N_1 + 1}^N$ for observations with $S_i = 0$
(henceforth the {\it external} sample $\Sett{S}_0$). The {\it overall
} sample is $\Sett{S} := \Sett{S}_1 \cup \Sett{S}_0$.

\vspace{-3mm}
\paragraph*{Estimands}

We write counterfactual means as $\phi = \Exp{Y^{a, S = 1}}$ for generalizability and $\Exp{Y^a | S = 0}$ for transportability, and contrasts between such counterfactual means under any two treatment levels $a, a'$ represent the average treatment effects (ATE). `Standard' estimation of effects in the study sample under unconfoundedness is a well-studied and largely resolved problem (see \parencite{Kennedy2022-jj} for a review). We study the generalization and transportation problems in the present paper. To this end, we make the following assumptions:

\be[noitemsep,topsep=0pt]
\I Consistency / SUTVA : $Y_i = \Indic{A_i = a} Y^a_i$ 
\I Ignorability of Treatment: $Y^0, \dots, Y^a \indep A | \ve{X = x}, S = 1$ 
\I Overlap
    \be
    \I Treatment overlap: $0 < \Prob{A = a | \vee{X = x} , S = 1} < 1$
    \I Selection overlap: $0 < \Prob{S = 1 | \vee{X = x}} < 1$
    \ee
\I Selection 
    \be
        \I $Y^0, \dots, Y^a \indep S | \vee{X = x}$  Ignorability of Selection. 
        \I $\Exp{Y|A, \vee{X}, S = 1} = \Exp{Y|A, \vee{X}, S = 0}$. The outcome model is stable across S strata.
    \ee
\ee

Under assumptions 1,2,3 and 4a, causal quantities of interest in the overall sample are identified \parencite{Bia2020-lu,Dahabreh2019-cr}, while under 1,2,3, and 4b, causal quantities of interest are identified in the target sample \parencite{Dahabreh2020-ys}. While prior work focused on binary treatments, we establish identification and estimation for counterfactual means and causal contrasts for multiple discrete treatments, which is the norm at Netflix and other industry settings.

% \vspace{-10mm}
\paragraph*{Estimators}

Our preferred estimators are efficient influence function (EIF) based that take the form of sample averages $\wh{\psi} = \ooN \sumin \varphi(W_i)$ where $W_i = (A_i, \vee{X}_i, Y_i, S_i)$. The influence function obeys  $n^{1/2} (\wh{\psi} - \psi) = n^{-1/2} \sumin \varphi(W_i) + o_p(1)$. This form characterizes Regular and Asymptotically Linear (RAL) estimators and allows us to construct valid confidence intervals using the sample variance of the influence function. These estimators rely on the estimation of three nuisance functions: (1) Outcome model $\mu^a(\vee{x}) = \Exp{Y | A = a , \vee{X = x}}$, (2) Treatment Propensity score $\pi^a(\vee{x}) = \Prob{A = a | \vee{X = x} , S = 1}$, and (3) Selection propensity score $\rho(x) = \Prob{S = 1 | \vee{X = x}}$. Their sample analogues $\wh{\alpha}$ are fit using machine learning estimators with cross-fitting and is implemented in the package with regularized regressions and generalized random forests. 

% The former scales to large datasets better than the latter.

The estimators under consideration take on one of three forms outlined in table~\ref{tab:table-estimators}. \textbf{Outcome Modeling (OM)} is a pure transfer-learning approach that involves fitting conditional response surfaces $\Exp{Y | A = a, S = 1}$ over the observations with nonmissing $Y$ and extrapolating these over the relevant samples.  \textbf{Inverse Selection Weighting (ISW)} involves modeling the selection probability into the source sample with covariates, and reweighting observations to mimic the target samples.
\textbf{Augmented ISW (AISW)} combines the ISW and OM approaches by augmenting the outcome model with an weighted average of residuals ($Y - \mu(\cdot)$) and possesses double-robustness properties (from outcome of both propensity models) analogous to the classical Augmented IPW estimator \parencite{robins1994estimation}. Both ISW and AISW can be stabilized using a Hajek normalization term equal to the sum of weights in each treatment level $a$.

If only summary statistics are available for the target sample, AISW is infeasible since it requires individual level covariates for all observations to construct weights. In such cases, a `calibration' approach based on solving for balancing weights that ensure balance between two population is feasible and has appealing properties in finite samples , and is also implemented using entropy loss \parencite{Hainmueller2012-rd} in the package.

\begin{table}[H]
% \fns
\centering 
\caption{\label{tab:table-estimators} Estimators. Difference between marginal means $\wh{\psi}_a - \wh{\psi}_{a'}$ yields causal contrasts $\tau(a, a')$. Standard errors are computed as $\sqrt{\hat{\sigma}^2/n}$ where $\wh{\sigma^2}$ is the sample variance of the influence function of interest (marginal mean or causal contrast) for AISW, or via the nonparametric or bayesian bootstrap for other estimators}

\centerline{
\begin{tabular}{|l | L | L |}
\hline
& 
    \makecell{
        \textbf{Generalization} \\
        \sum_x \Exp{Y | A = a, S = 1, \vee{X}} P(\vee{X})
    }
         & 
    \makecell{
        \textbf{Transportation}  \\ 
        \sum_x \Exp{Y | A=a, S = 1,\vee{X}} P(\vee{X}| S = 0)
    }
        \\
\hline
OM & 
     \frac{1}{n}  \sum_{i}  \wh{\mu}^{a}(\vee{X}_i) &
     \frac{1}{\abs{\Sett{S}_0}} \sum_i (1 - S_i) \wh{\mu}^{a}(\vee{X}_i)  \\
ISW &
    \frac{1}{n}  \sum_{i}     \frac{S_i}{\wh{\rho}(\vee{X}_i)}  
        \frac{\Indic{A = a}}{\wh{\pi}^a(\vee{X}_i)} Y_i &
    \frac{1}{n} 
    \sum_i 
    \frac{1}{\wh{\E}[S_i = 0]} 
    \frac{S_i(1 - \wh{\rho}(\vee{X}_i))} { \wh{\rho}(\vee{X}_i)}  
        \frac{\Indic{A = a}}{\wh{\pi}^a(\vee{X}_i)} Y_i \\
AISW &
    \frac{1}{n}  \sum_{i} 
    \wh{\mu}^{a}(\vee{X}_i) +
        \frac{S_i}{\wh{\rho}(\vee{X}_i)}
        \frac{\Indic{A = a}}{\wh{\pi}^a(\vee{X}_i)} (Y_i - \wh{\mu}^a(\vee{X}_i) ) &
    % \frac{1}{n}
    \frac{1}{n}  \sum_{i}       
    \frac{1}{\wh{\E}[S_i = 0]}
    \Bigpar{
      (1-S_i) \wh{\mu}^{a}(\vee{X}_i) +
       \frac{S_i (1 - \wh{\rho}(\vee{X}_i))}{\wh{\rho}(\vee{X}_i)}
          \frac{\Indic{A = a}}{\wh{\pi}^a(\vee{X}_i)} (Y_i - \wh{\mu}^a(\vee{X}_i) )
    }\\
\hline
\end{tabular}
}

\end{table}

\section*{Simulation Study}

We study the generalization estimators' performance in a simulation study, where we simulate four scenarios where covariates $x_1, \dots, x_{10} \sim \mathsf{U}[-1, 1]$ and treatment is randomly assigned with probability $0.5$, and the true (selection / outcome) models are (linear / nonlinear), and nuisance functions are estimated using regularized 
linear regressions with $\lambda$ set to minimise CV-MSE. The selection model dictates how the study sample is selected from the target population, and whenever this is a function of covariates, the experimental estimate of the average treatment effect (SATE) is biased for the treatment effect in the target population (PATE).

\begin{figure}[htbp]
\caption{Simulation Results}
\label{fig:simfig}
% MSE
\subfloat[RMSE]{
  \includegraphics[width=.5\textwidth]{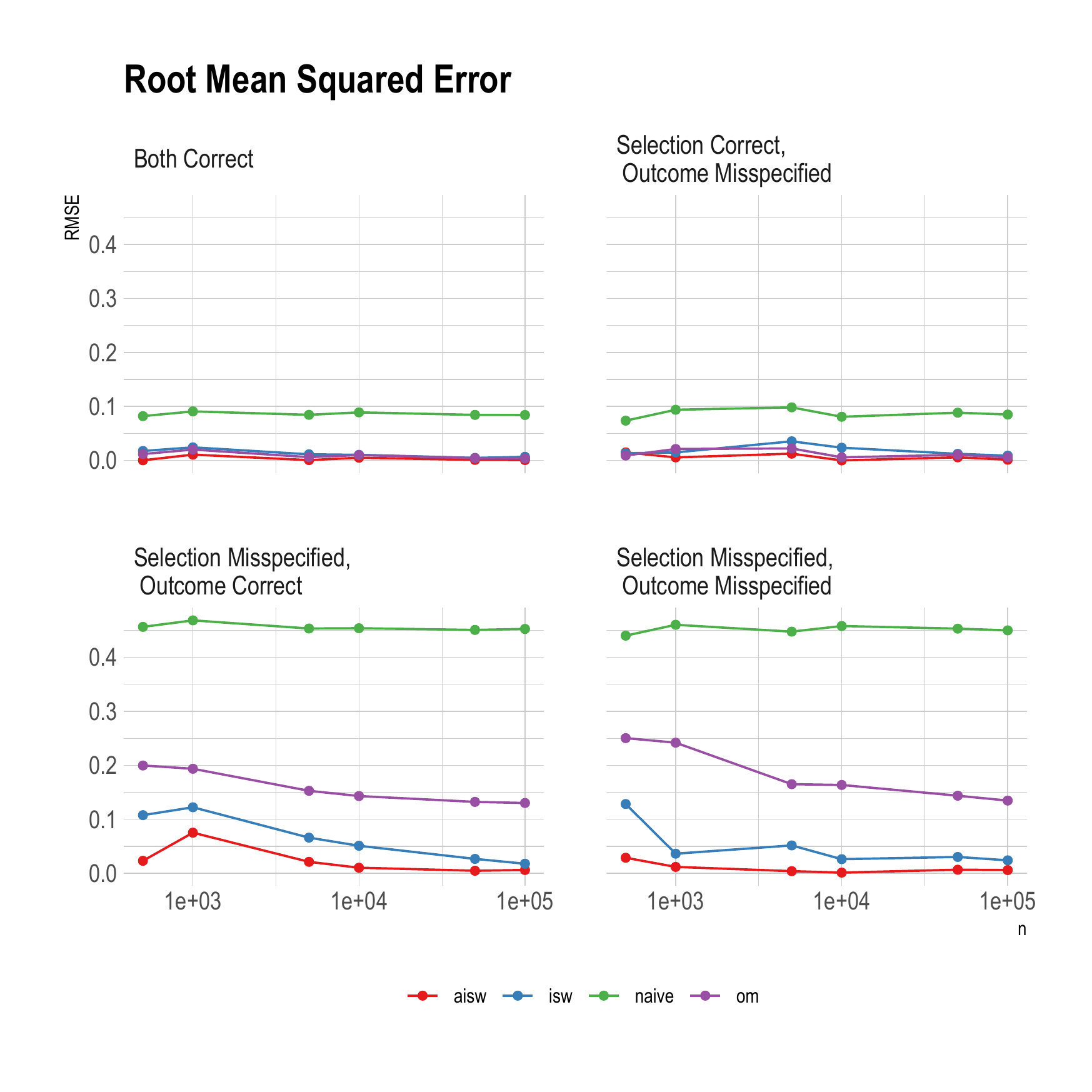}
}  
% bias
\subfloat[Bias]{
  \includegraphics[width=.5\linewidth]{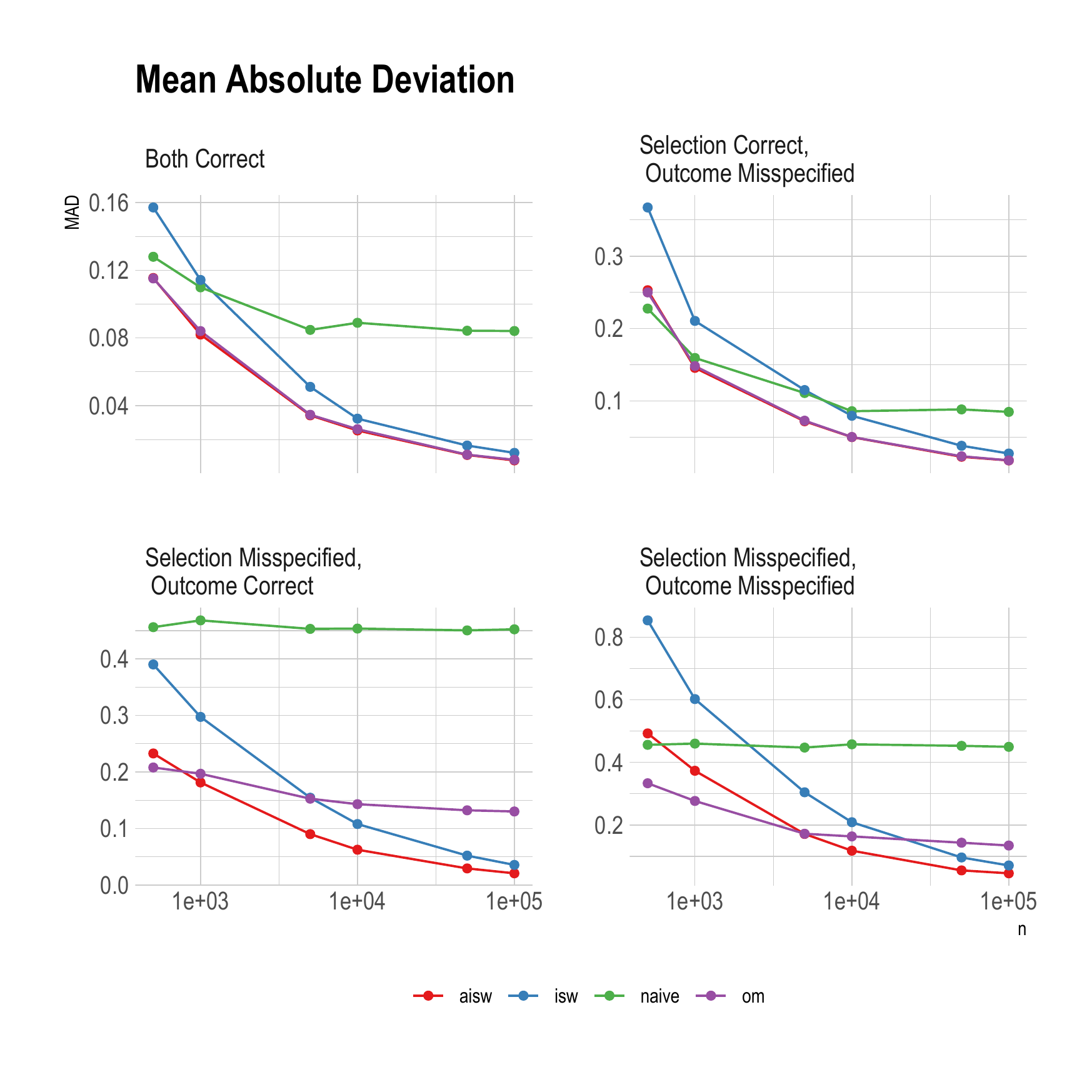}
} 
\\
\subfloat[95\% CI coverage]{
% coverage
  \includegraphics[width=.5\linewidth]{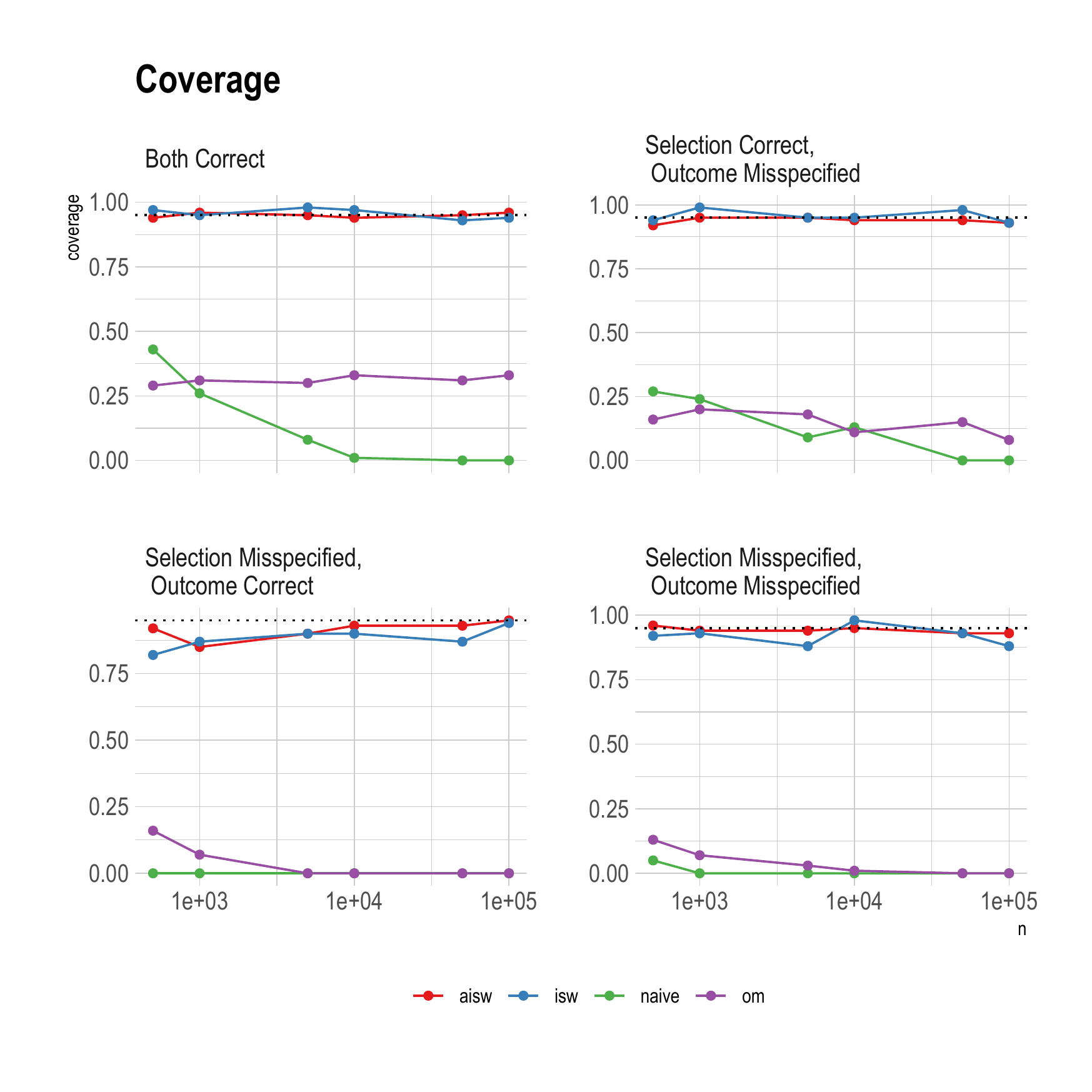}
} 
\subfloat[Runtime]{
  \includegraphics[width=.5\linewidth]{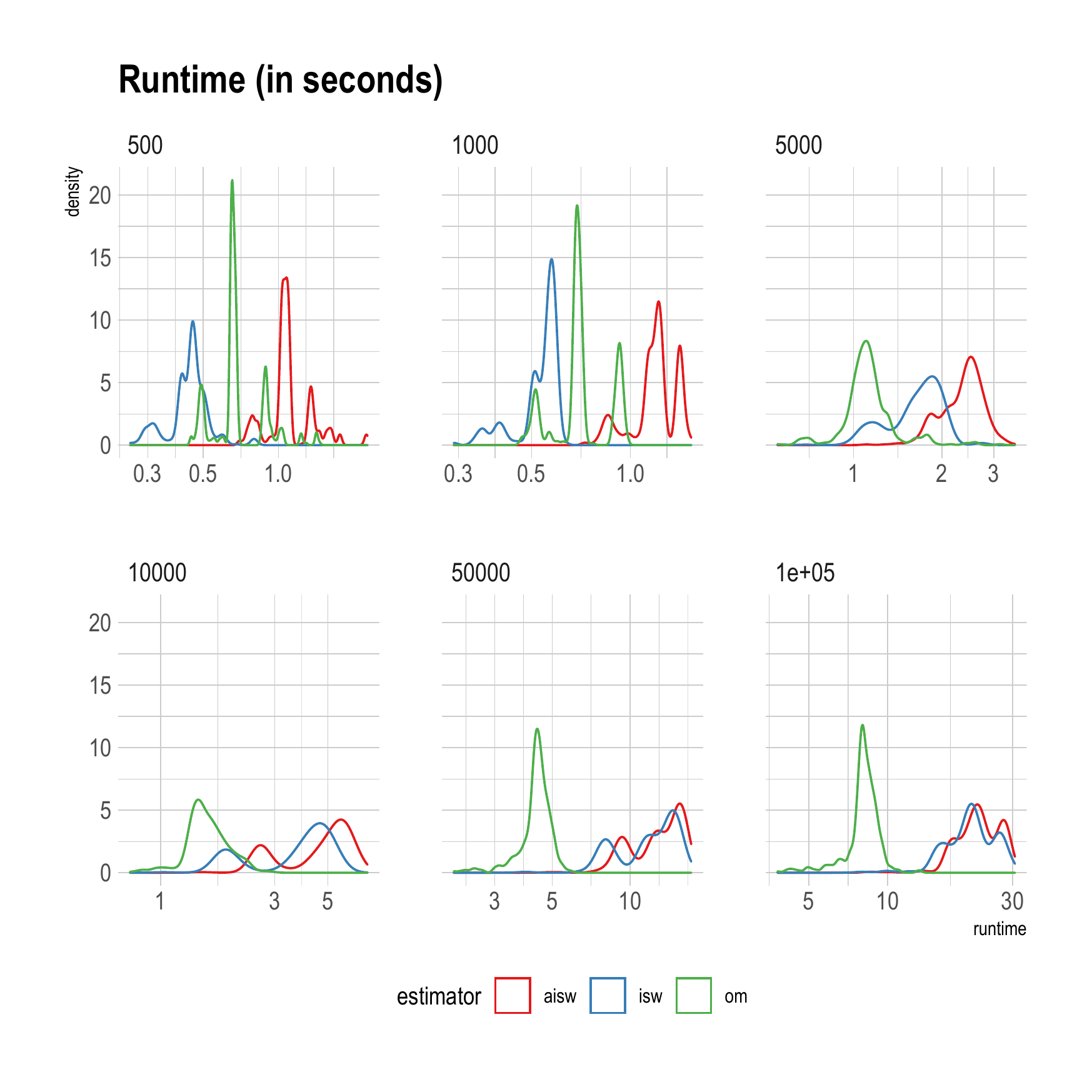}
}
\end{figure}

We report the performance of the above estimators in figure~\ref{fig:simfig}, which displays the RMSE, Bias, Coverage rate, and runtime across 500 replications. We find that reweighting estimators (red, blue, and purple) consistently outperform `naive' SATE estimates (green). Consistent with the analogous results in the unconfoundedness literature, we find that among the generalization estimators, the augmented inverse selection weighting estimator (red) performs best along MSE, bias, and variance dimensions. 

\section*{Discussion}

In this work, we provide a concise multi-treatment framework for
causal generalization and transportation, and provide a performant
computational implementation for it. In current practice,
practitioners often informally perform `naive' extrapolation of the
Sample Average Treatment Effect (SATE) to the Target Average Treatment
Effect (TATE), which may result in erroneous conclusions arising from
three distinct sources of bias generated by differences between study
and external samples : (1) unequal distribution of effect modifier
covariates in the two samples, (2) Lack of covariate overlap between
the two samples, and (3) differences between the effect modification
functions between the two samples\footnote{We provide a fuller
derivation of these in appendix~\ref{appdx:biasDecomp}}.

Naive extrapolation can be improved upon by using the framework
presented in the present paper, which shows that causal generalization
and transportation involves (1) estimating strata-specific Conditional
Average Treatment Effects (CATEs) $\tau_s(\vee{X})$, (2) asserting
outcome model stability wherein the heterogeneity function
$\tau(\vee{X})$ is stable across the study and external sample, and
(3) reweighting CATEs to match the covariate distribution $p(\vee{X})$
in the target population. The feasibility of each of these steps may
be problem-specific, and practitioners are advised to carefully
consider potential problems in each step in their particular
application.

While an enormous literature has emerged on CATE estimation (see
\parencite{Knaus2020-ah} for a review), evaluating the quality of the
resultant estimates remains a challenging problem. Omnibus tests for
systematic treatment effect heterogeneity
\parencite{Ding2019-nr}(implemented as \texttt{dfmTest} in the
package) and \parencite{chernozhukov2020generic} are strongly
recommended prior to the use of generalization estimators. CATE
estimation for settings with small treatment effects as is typical in
industry settings is a growing area of research
\parencite{athey2021semiparametric} and progress on this problem can
readily be integrated to improve upon effect transportation.

Outcome model stability across the study and external samples is
inherently untestable, and must be justified from substantive
knowledge of the study and target population, as well as the nature of
the experimental manipulation. For example, a video compression
treatment might have systematic and stable treatment effects across
subpopulations, while a recommendation algorithm may not because of
preference heterogeneity that isn't adequately captured by covariates.
Sensitivity analyses in the vein of
\parencite{chernozhukov2022long,dorn2022sharp,Nie2021-ku} that assess
the magnitude of violations of outcome model stability that overturn
transportation conclusions are a fruitful avenue for future research.

Finally, the effectiveness of selection weights in reducing the
imbalance between the study and external populations can be evaluated
using the suite of tools developed for propensity score weights.
Plotting standardized mean differences between the source and target
samples is a reasonable first check for whether the weights are
effective at reducing imbalance, and is implemented as the plot method
for the model object in the package. These figures are also also
intended to help choose between `indirect' balancing via propensity
score modelling versus calibration approaches that target balance
directly.

%%%%%%%%%%%%%%%%%%%%%%%%%%%%%%%%%%%%%%%%%%%%%%%%%%%%%%%%%%%%
% \pagebreak
%%%%%%%%%%%%%%%%%%%%%%%%%%%%%%%%%%%%%%%%%%%%%%%%%%%%%%%%%%%%
% APPENDIX
%%%%%%%%%%%%%%%%%%%%%%%%%%%%%%%%%%%%%%%%%%%%%%%%%%%%%%%%%%%%
% \newpage
\renewcommand{\thetable}{A\arabic{table}}
\renewcommand{\thefigure}{A\arabic{figure}}
\setcounter{table}{0}
\setcounter{figure}{0}

% section simulations (end)
% \printbibliography
\renewcommand{\mkbibnamefamily}[1]{\textsc{#1}}
\printbibliography

@ARTICLE{Dahabreh2019-cr,
  title    = "Generalizing causal inferences from individuals in randomized
              trials to all trial-eligible individuals",
  author   = "Dahabreh, Issa J and Robertson, Sarah E and Tchetgen, Eric J and
              Stuart, Elizabeth A and Hern{\'a}n, Miguel A",
  journal  = "Biometrics",
  volume   =  75,
  number   =  2,
  pages    = "685--694",
  month    =  jun,
  year     =  2019,
  language = "en"
}

@ARTICLE{Dahabreh2020-ys,
  title    = "Extending inferences from a randomized trial to a new target
              population",
  author   = "Dahabreh, Issa J and Robertson, Sarah E and Steingrimsson, Jon A
              and Stuart, Elizabeth A and Hern{\'a}n, Miguel A",
  journal  = "Statistics in medicine",
  volume   =  39,
  number   =  14,
  pages    = "1999--2014",
  month    =  jun,
  year     =  2020,
  language = "en"
}

@ARTICLE{Nie2021-ku,
  title         = "Covariate Balancing Sensitivity Analysis for Extrapolating
                   Randomized Trials across Locations",
  author        = "Nie, Xinkun and Imbens, Guido and Wager, Stefan",
  month         =  dec,
  year          =  2021,
  url           = "http://arxiv.org/abs/2112.04723",
  archivePrefix = "arXiv",
  primaryClass  = "econ.EM",
  eprint        = "2112.04723"
}

@ARTICLE{Hainmueller2012-rd,
  title     = "Entropy Balancing for Causal Effects: A Multivariate Reweighting
               Method to Produce Balanced Samples in Observational Studies",
  author    = "Hainmueller, Jens",
  journal   = "Political analysis: an annual publication of the Methodology
               Section of the American Political Science Association",
  publisher = "Cambridge University Press",
  volume    =  20,
  number    =  1,
  pages     = "25--46",
  year      =  2012
}

@ARTICLE{Bia2020-lu,
  title         = "Double machine learning for sample selection models",
  author        = "Bia, Michela and Huber, Martin and Laff{\'e}rs, Luk{\'a}{\v
                   s}",
  month         =  nov,
  year          =  2020,
  url           = "http://arxiv.org/abs/2012.00745",
  archivePrefix = "arXiv",
  primaryClass  = "econ.EM",
  eprint        = "2012.00745"
}

@ARTICLE{Kennedy2022-jj,
  title         = "Semiparametric doubly robust targeted double machine
                   learning: a review",
  author        = "Kennedy, Edward H",
  month         =  mar,
  year          =  2022,
  url           = "http://arxiv.org/abs/2203.06469",
  archivePrefix = "arXiv",
  primaryClass  = "stat.ME",
  eprint        = "2203.06469"
}

@ARTICLE{Knaus2020-ah,
  title     = "Machine learning estimation of heterogeneous causal effects:
               Empirical Monte Carlo evidence",
  author    = "Knaus, Michael C and Lechner, Michael and Strittmatter, Anthony",
  journal   = "The econometrics journal",
  publisher = "Oxford Academic",
  volume    =  24,
  number    =  1,
  pages     = "134--161",
  month     =  jun,
  year      =  2020,
  language  = "en"
}

@techreport{chernozhukov2022long,
  title={Long story short: Omitted variable bias in causal machine learning},
  author={Chernozhukov, Victor and Cinelli, Carlos and Newey, Whitney and Sharma, Amit and Syrgkanis, Vasilis},
  year={2022},
  institution={National Bureau of Economic Research}
}

@article{dorn2022sharp,
  title={Sharp sensitivity analysis for inverse propensity weighting via quantile balancing},
  author={Dorn, Jacob and Guo, Kevin},
  journal={Journal of the American Statistical Association},
  number={just-accepted},
  pages={1--28},
  year={2022},
  publisher={Taylor \& Francis}
}

@InProceedings{pmlr-v130-phan21a,
  title = 	 { Designing Transportable Experiments Under S-admissability },
  author =       {Phan, My and Arbour, David and Dimmery, Drew and Rao, Anup},
  booktitle = 	 {Proceedings of The 24th International Conference on Artificial Intelligence and Statistics},
  pages = 	 {2539--2547},
  year = 	 {2021}
}

@techreport{athey2021semiparametric,
  title={Semiparametric Estimation of Treatment Effects in Randomized Experiments},
  author={Athey, Susan and Bickel, Peter J and Chen, Aiyou and Imbens, Guido and Pollmann, Michael},
  year={2021},
  institution={National Bureau of Economic Research}
}

@misc{chernozhukov2020generic,
      title={Generic Machine Learning Inference on Heterogenous Treatment Effects in Randomized Experiments}, 
      author={Victor Chernozhukov and Mert Demirer and Esther Duflo and Iván Fernández-Val},
      year={2020},
      eprint={1712.04802},
      archivePrefix={arXiv},
      primaryClass={stat.ML}
}

@ARTICLE{Ding2019-nr,
  title     = "Decomposing Treatment Effect Variation",
  author    = "Ding, Peng and Feller, Avi and Miratrix, Luke",
  journal   = "Journal of the American Statistical Association",
  publisher = "Taylor \& Francis",
  volume    =  114,
  number    =  525,
  pages     = "304--317",
  month     =  jan,
  year      =  2019,
}

@article{robins1994estimation,
  title={Estimation of regression coefficients when some regressors are not always observed},
  author={Robins, James M and Rotnitzky, Andrea and Zhao, Lue Ping},
  journal={Journal of the American statistical Association},
  volume={89},
  number={427},
  pages={846--866},
  year={1994},
  publisher={Taylor \& Francis}
}

\appendix
\part{Appendix}

\section*{Bias decomposition}\label{appdx:biasDecomp}

To simplify notation, lets assume covariate $\vee{X}_i \in \mathcal{X}$ is discrete, and follows distributions $p_s(\vee{X})$ and $p_t(\vee{X})$ in the study and target samples respectively. We denote the Conditional Average Treatment Effect (CATE) as $\tau_{k}(\vee{x}), k \in \{s, t\}$. We assume CATE stability across samples $\tau_s(\vee{x}) = \tau_t(\vee{x}) = \tau(\vee{x})$, and discuss implications of relaxing this assumption next.

The gap between the Target Average Treatment Effect (TATE) and Sample Average Treatment Effect (SATE) can be decomposed as follows

\begin{align*}
\text{TATE - SATE} & = \sum_{x \in \Sett{X}_t} p_t(\vee{x}) \tau_t(\vee{x}) - p_s(\vee{x}) \tau_s(\vee{x})  \\
 & = \sum_{x \in \Sett{X}_t} \Bigpar{p_t(\vee{x}) - p_s(\vee{x})} \tau(\vee{x}) & \text{by } \tau_s(\cdot) = \tau_t(\cdot) = \tau(\cdot) \\
 & = \sum_{x \in \Sett{X}_t} p_s(\vee{x}) \Bigpar{\frac{p_t(\vee{x})}{p_s(\vee{x})} - 1} \tau(\vee{x})
\end{align*}

From the above, we can see three distinct sources of bias:
\be
\I When overlap holds, bias contributions come from strata where the following three conditions are true
    \be
    \I $\tau(\vee{x}) \neq 0$: Non-zero treatment effects
    \I $p_s(\vee{x}) > 0$ : Nonzero support in study population
    \I $p_t(\vee{x}) \neq  p_s(\vee{x})$ : Distribution of covariate $\vee{x}$ is different across study and target population
    \ee
\I  When overlap is violated such that $p_t(\vee{x}) > 0 \AND p_s(\vee{x})$: there exist strata in the target population that are unrepresented in the study, the SATE is biased and the bias is increasing in the size of $p_t(\vee{x})$ (fraction of target sample unrepresented in study population) and $\tau(\vee{x})$
\I If the CATE functions $\tau_s(\vee{x}) \neq \tau_s(\vee{x})$ (i.e. effect modification is different between the study and target sample).
\ee

Our reweighting approach addresses (1) using a post-stratification weights for discrete covariates and its analogue selection score for continuous covariates. (2) and (3) produce bias that is impossible to resolve without additional  data collection (for 2, for example through the use of S-admissable designs \parencite{pmlr-v130-phan21a}) or prior knowledge of CATE functions (for 3).
\medskip

\end{document}